\documentclass{optica-article}

\journal{opticajournal} 

\articletype{Research Article}

\usepackage{lineno}

\newcommand{\ket}[1]{\lvert #1 \rangle}

\begin{document}

\title{Finite key analysis of experimentally realized practical COW-QKD protocol}

\author{Neha Pathania,\authormark{1,2,*} Sandeep Mishra,\authormark{1,+} and Anirban Pathak\authormark{1,@}}

\address{\authormark{1} Department of Physics and Materials Science and Engineering, Jaypee Institute of Information Technology, A 10, Sector 62, Noida, UP-201309, India\\\authormark{2} Department of Electronics and Communication Engineering, Indian Institute of Technology Roorkee, India\\
}

\email{\authormark{*}neha1.ece@sric.iitr.ac.in, \authormark{+}sandeep.mtec@gmail.com, \authormark{@}anirban.pathak@jiit.ac.in} 


\begin{abstract*} 
An experimental implementation of the Coherent One-Way Quantum Key Distribution (COW-QKD) protocol is reported under realistic conditions, and a clean and easy-to-use framework for performing finite key analysis of the COW-QKD protocol is provided by extending a set of existing results. The framework provided here is used to perform finite key rate analysis of the COW-QKD protocol with respect to the actual parameters used in the experimental realization reported here. The system is kept running for several hours with different experimental parameters and stable secure key rates between 1.2 to 1.6 kbps are observed.  In addition, QBER, phase error rate and secure key rate are obtained under finite key analysis, and it is shown that COW-QKD is secure for medium-range transmissions (up-to $\sim156$ $(171)$ km of optical fiber with 0.2 dB loss per km if detector efficiency is 0.1 (0.2) and other parameters are same as those used in this experiment). 
\end{abstract*}

\section{Introduction}
 The advent of large-scale quantum computing poses a fundamental threat to classical public-key cryptographic systems, accelerating the need for cryptographic primitives whose security is guaranteed by the laws of quantum mechanics rather than computational hardness \cite{steane1998quantum,o2007optical,arute2019quantum,pirandola2020advances,portmann2022security}. In parallel, rapid progress in photonic quantum information processing has established photons as carriers of quantum information, owing to their long coherence times, low environmental coupling, and seamless compatibility with optical fiber and integrated photonic technologies \cite{slussarenko2019photonic,psiquantum2025,romero2024photonic}. These advantages make photonic platforms particularly well-suited for quantum communication in general and quantum cryptography in particular. In fact, a qubit can be realized in different two-level quantum systems, but only photonic qubits can be transmitted easily from one place to another. Now, even in photonic implementations, quantum information can be encoded in different degrees of freedom of light, including polarization, phase, time bins, and coherent-state amplitudes \cite{beveratos2002single,Brendel1999,Weedbrook2012,Gisin2002,Scarani2009,xu2020secure}. Among these, time-bin and phase-encoded photonic qubits are especially robust against polarization drift and fiber birefringence, enabling stable transmission over long distances in standard telecommunication fiber. As a result, photonic technologies have recently been significantly developed to realize different protocols for practical quantum key distribution (QKD). In fact,  several protocols for have been demonstrated in both laboratory and field environments \cite{o2009photonic,aldama2022integrated,pitkanen2011efficient,xu2020secure}.

 QKD enables two distant users to establish a shared secret key with information-theoretic security by exploiting fundamental quantum principles such as measurement leading to disturbance and the impossibility of perfectly distinguishing non-orthogonal states. Over the years, a wide range of QKD protocols have been proposed \cite{xu2020secure}, including discrete-variable schemes based on single-photon detection \cite{bennett2014quantum,djordjevic2025discrete} and continuous-variable protocols relying on homodyne or heterodyne measurements \cite{zhang2024continuous}. While early protocols were aimed at providing strong theoretical security guarantees, their practical deployment was often constrained by source imperfections, detector inefficiencies, and limited key generation rates \cite{pirandola2020advances}. To address these challenges, distributed phase reference (DPR) protocols, such as differential phase shift (DPS) \cite{inoue2002differential} and coherent one-way (COW) \cite{stucki2005fast} QKD protocols were introduced. These protocols employ trains of weak coherent pulses and encode information in the relative phase or temporal structure between successive pulses, allowing operation at high repetition rates with comparatively simple designs. Importantly, DPR protocols are naturally compatible with fiber-based interferometric detection schemes and have demonstrated favourable performance in terms of achievable distance and key rate \cite{wang20122,shibata2014quantum,korzh2015provably,malpani2024implementation,kumar2024experimental}. Among DPR schemes, COW-QKD has attracted particular interest due to its experimental simplicity and robustness. In COW-QKD, logical bits are encoded in the presence or absence of coherent pulses across adjacent time bins, while security is monitored by measuring the coherence between successive pulses using an unbalanced Mach–Zehnder interferometer \cite{stucki2005fast}. The performance of a protocol critically depends on optical parameters such as interferometer visibility, phase stability, fiber-induced birefringence, and detector characteristics, including efficiency, dark count rate, and dead time \cite{malpani2024implementation,kumar2024experimental}. Fluctuations in these parameters directly affect the quantum bit error rate (QBER) and hence affect the secure key rates.

Furthermore, in contrast to the ideal situation where asymptotically large key size is usually assumed for obtaining the security proof for the QKD protocols, the practical QKD systems inevitably operate in the finite-key regime, where security parameters must be estimated from finite data blocks \cite{scarani2008quantum,scarani2008security}. Finite-key effects introduce statistical uncertainties that reduce the extractable secret key rate compared to asymptotic predictions \cite{Scarani2009}. For COW-QKD, finite-key analysis is especially nontrivial due to the reliance on coherence measurements and correlated detection events, making rigorous parameter estimation essential for realistic security claims. Environmental instabilities further amplify statistical fluctuations, underscoring the need for experimental studies that explicitly combine optical performance analysis with finite-key security evaluation. Here, it may be noted that the recent studies have shown that the current COW-QKD system is insecure against coherent attack \cite{branciard2008upper,trenyi2021zero} and can only distribute secret keys safely within 20 km of the optical fiber length \cite{gonzalez2020upper}. However, in 2022, Gao et al. proposed a practical implementation of COW-QKD by adding a two-pulse vacuum state as a new decoy sequence \cite{gao_2022}. This proposal maintains the original experimental setup as well as the simplicity of its implementation. Utilizing detailed observations on the monitoring line to provide an analytical upper bound on the phase error rate, they provide a high-performance COW-QKD asymptotically secure against coherent attacks.  However, there was still a problem with its practical implementation. As mentioned above, generally, the security proofs are obtained using the asymptotically large key size, but real physical implementations have only a finite amount of resources. So, finite key analysis of realistic implementation of QKD systems is the need of the hour. Continuing from this, Li et al. \cite{li2024finite} worked towards the finite-key security analysis for the variant of COW-QKD also includes two-pulse vacuum state. So, we can see that this variant of COW-QKD is secure against zero error attack and this can be used for secure transmission with good key rates. Taking  a step forward we have implemented two-vacuum pulse variant of COW-QKD in our laboratory and undertook the tight finite key analysis of the system under realistic environmental conditions. We have shown that if we consider the statistics of QBER under realistic environmental conditions with a telecom-grade optical fiber and SPAD, the distance of COW-QKD can be enhanced to about 175 km. Further, if we take into account the estimation of phase error, secure key rates can be generated over distance of about 90 km.  Hence, from the realistic implementation of COW-QKD, we can say that the system is secure under tight finite key regime for a medium range of distance.

The rest of the paper is structured as follows. A brief overview of COW-QKD protocol with emphasis on its two vacuum pulse variant is provided in Section \ref{COW}. This is followed by the finite key analysis of the COW-QKD protocol in Section \ref{finite}, where a general and easy-to-use framework for performing finite key analysis of COW-QKD protocol is developed along the lines of ~\cite{gao_2022,li2024finite} without being specific to the experimental parameters relevant to the present work. Subsequently, in Section \ref{experiment}, the framework provided in the previous section is used to perform finite key analysis of the COW-QKD protocol implemented in the present work. Specifically, finite key analysis framework is used for our experimentally realized system to obtain the upper bounds on quantum bit error rate (QBER), phase error rate and secure key rate. Finally, the paper is concluded in Section \ref{conclusion}.

\section{COW-QKD Protocol Description} \label{COW}

 Owing to its conceptual simplicity, robustness against polarization fluctuations, and compatibility with standard telecommunication components, COW-QKD \cite{stucki2005fast} has emerged as a promising candidate for practical long-distance quantum key distribution. Further, the security is derived from the phase coherence between successive optical pulses rather than from single-photon encoding. Here, in this section, we outline the COW-QKD protocol and highlight the key operational steps required for the analysis presented in this work. 
\subsection{State Preparation and Encoding}
In the COW-QKD protocol, the sender (Alice) employs a continuous-wave laser source followed by an intensity modulator to generate a sequence of weak coherent pulses at a fixed repetition rate. Logical bits are encoded across two consecutive time bins using the presence or absence of optical pulses. Specifically, the two-mode state for bit values $0$ and $1$ are defined as
\begin{equation}
\begin{aligned}
0 &\equiv \ket{\alpha}_{2j-1}\ket{0}_{2j}, \\
1 &\equiv \ket{0}_{2j-1}\ket{\alpha}_{2j},
\end{aligned}
\end{equation}
where $\ket{\alpha}$ denotes a coherent state with mean photon number $\mu = |\alpha|^2 < 1$, and $\ket{0}$ represents the vacuum state. The states $\big|\alpha\rangle_{2j-1}\ket{0}_{2j}$ and $\ket{0}_{2j-1}\ket{\alpha}_{2j}$ are prepared with probabilities $\mathrm{p}_{z_1}$ and $\mathrm{p}_{z_2}$ respectively with $\mathrm{p}_{z_1}=\mathrm{p}_{z_2} $. In addition to the signal states, Alice randomly prepares two decoy states of the form $\ket{\alpha}_{2j-1}\ket{\alpha}_{2j}$ and $\big|0\rangle_{2j-1} \big|0\rangle_{2j}$ with probabilities $\mathrm{p}_{d_1}$ and, $\mathrm{p}_{d_2}$ respectively and $\mathrm{p}_{z_1}= \frac{1}{2} ( 1- \mathrm{p}_{d_1} -\mathrm{p}_{d_2})$. The objective of the decoy states is to monitor channel coherence and detect potential eavesdropping. In the two decoy state variants, this additional vacuum–vacuum decoy $\ket{0}_{2j-1}\ket{0}_{2j}$ is included to circumvent zero error attack. Each pair of time bins constitutes one protocol round, and Alice records the type of state sent in each round for later sifting and parameter estimation. The use of weak coherent states allows implementation without true single-photon sources while maintaining robustness against photon-number-splitting attacks.

\subsection{Measurement and Detection}

At the receiver side (Bob), the incoming optical signals are passively split using an asymmetric beam splitter with transmittance $t_B=0.90$, directing the majority of the signal to the data line (about $\approx 90\%$) and the remainder to the monitoring line. On the data line, Bob employs a single-photon detector ($D_T$) to record the arrival time of detection events, from which the raw key is generated based on the occupied time bin. The monitoring line consists of an unbalanced Mach–Zehnder interferometer with a one-bit delay, followed by two single-photon detectors ($D_{M_0}$ and $D_{M_1}$). This configuration enables Bob to test the phase coherence between successive non-empty pulses. Detection events occurring within well-defined interference time windows are used to estimate security-relevant parameters, while all other clicks are discarded. In practical implementations, all detectors are threshold detectors characterized by finite detection efficiency and non-zero dark count probability ($p_d$).

\subsection{Sifting and Parameter Estimation}

After the quantum transmission phase, Bob publicly announces the rounds in which detection events occurred on the data line. Alice retains only the corresponding logical bits to form the sifted key. Subsequently, Bob reveals the detection statistics obtained from the monitoring line, allowing Alice to estimate quantities such as the QBER on the data line and security parameters associated with coherence loss.

In recent variants of COW-QKD, the security analyses of COW-QKD and the monitoring statistics are used to bound the phase error rate rather than relying solely on interference visibility. This approach enables rigorous security proofs in both the asymptotic and finite-key regimes and allows composable security against general and coherent attacks ~\cite{gao_2022, li2024finite}.

\subsection{Classical Post-Processing}

If the estimated error rate lies below predefined thresholds, Alice and Bob proceed with classical post-processing. This includes error correction, during which a limited amount of information is disclosed over an authenticated classical channel, followed by privacy amplification to remove any information potentially available to an eavesdropper. The final secret key rate depends on experimental parameters such as detector efficiency, dead time, disclose rate, and compression ratio, all of which can be optimized while maintaining QBER within acceptable bounds.

\section{Finite Key Analysis} \label{finite}
\begin{figure}[htbp]
\centering\includegraphics[width=0.9\linewidth]{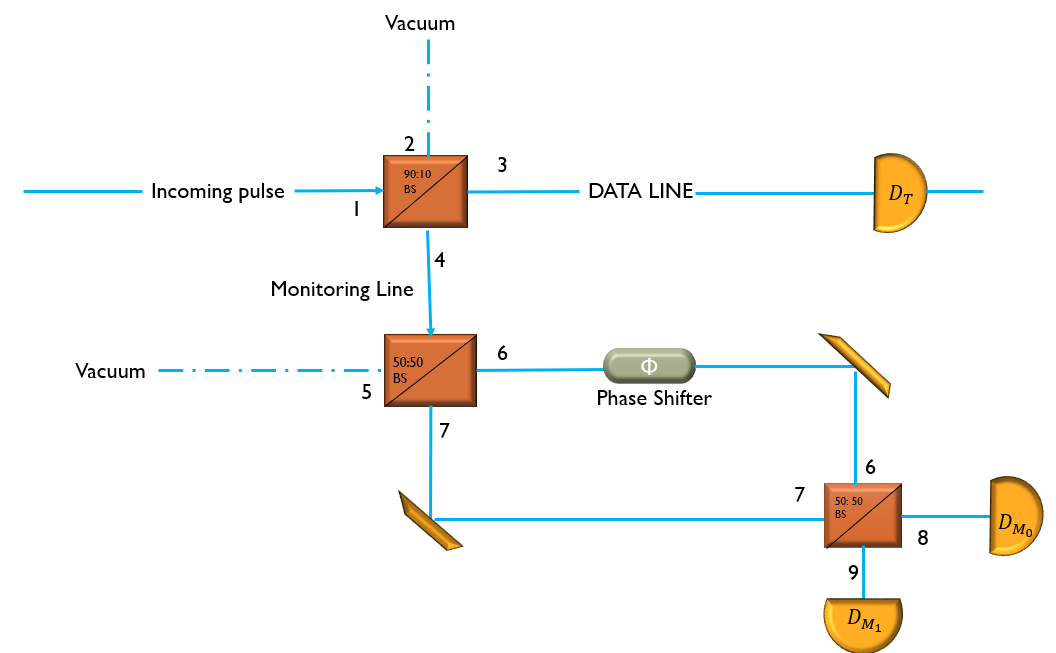}
\caption{(Color online) Schematic diagram of COW-QKD Protocol showing action of beam splitter}
\label{fig:beamsplitter}
\end{figure}

The COW-QKD setup (Fig.~\ref{fig:beamsplitter}) includes an asymmetric beam splitter (90:10) with transmittance ($t_B$), which splits the incoming pulses into the data line and the monitoring line. The data line includes a single photon avalanche detector (SPAD). The monitoring line consists of a  1-bit delay MZI setup with two SPAD detectors. On Alice's side we have a laser source with an attenuator to create a weak coherent pulse (WCP) with $\mu = 0.5$.  Alice randomly selects her initial bit string and encodes each bit into a two-pulse state. As the WCP passes through the asymmetric beam splitter, it either goes into the data line or the monitoring line. Bob's detection devices receive optical pulses transmitted through a quantum channel with transmittance of $\eta$.   On the data line, the quantum states are directly transmitted to the SPAD, which measures the arrival time of the pulses. On the monitoring line, Bob utilizes an MZI, which includes a one-bit delay and two single-photon detectors, to record which detector clicks within certain time windows. The detection efficiency of the detector $\eta_{d}$ and the dark-count rate probability  $p_d$ of each detector are assumed to be the same. 
The gains on the data line and the monitoring line are calculated by studying the action of the beam splitter on the WCPs, which are explained below.

\textbf{Action of Beam Splitter:}
The Beam Splitter (BS) operates in two-input, two-output modes. In Fig. \ref{fig:beamsplitter}, we have 1 and 2 as input modes while 3 and 4 are output modes. Reflection from the beam splitter will introduce a phase $\iota$. The incoming pulse will come through mode 1 and in quantum optics we generally consider a vacuum from mode 2 of BS. The output from mode 3 will go into the data line and the output from mode 4 will be fed into the monitoring line.\\
The gains on the data line are : $G_{0_z}^{\tau_0} , G_{0_z}^{\tau_1} ,  G_{1_z}^{\tau_0} ,  G_{1_z}^{\tau_1}$ where, $G_{0_z}^{\tau_0}$ denotes the gain on the data line for the state $\big|0_z\rangle = \big|\alpha\rangle \big|0\rangle$ at $\tau_0 = 2j-1$ ;  $G_{0_z}^{\tau_1}$ denotes the gain on the data line for the state $\big|0_z\rangle = \big|\alpha\rangle \big|0\rangle$ at $\tau_1 = 2j$, with $\big|\alpha\rangle$ being the coherent state. Similarly, $G_{1_z}^{\tau_0}$ denotes the gain on the data line for the state $\big|1_z\rangle = \big|0\rangle \big|\alpha\rangle$ at $\tau_0 = 2j-1$  and $G_{1_z}^{\tau_1}$ denotes the gain on the data line for the state $\big|1_z\rangle = \big|0\rangle \big|\alpha\rangle$ at $\tau_1 = 2j$.
The gain of the incoming state $\varphi$ heralded by the detector $D_{M_{i}}$ ($i=0$ or $1$) is calculated using the expression:
\begin{equation}
    G_{\varphi}^{M_i} = \langle \varphi \big| \hat{K}_i^\dagger \hat{K}_i \big| \varphi \rangle,
    \label{gaineqn}
\end{equation}
where $\hat{K}_i$ is the Kraus operator corresponding to detector $D_{M_i}$. In the modeling of practical imperfections, we consider the role of the detector as an amplitude-damping channel, which captures loss mechanisms during detection. 
To calculate click probability/gain, the expression becomes $ K_{0}= 1 - \big|0 \rangle \langle 0 \big|$ for at least one photon click and $K_{1}= \big|0 \rangle \langle 0 \big|$ for no photon click.
The coherent state $\big|\alpha \rangle $ in the fock basis is written as : 
\begin{align*}
    \big|\alpha \rangle = e^{-|\alpha|^2/2} \sum_{n=0}^{\infty} \frac{\alpha^n}{\sqrt{n!}} \big|n\rangle.
\end{align*}
In the beam splitter, the input and output modes are related according to, 
\begin{equation}
    \hat{a}_3 = \sqrt{ \eta t_B} \hat{a}_1 + \iota \sqrt{\eta (1- t_B)} \hat{a}_2, 
    \end{equation} and 
    \begin{equation}
        \hat{a}_4 = \iota \sqrt{\eta (1-t_{B})} \hat{a}_1 + \sqrt{\eta t_B} \hat{a}_2.
    \end{equation}
Expressing $\hat{a}_1$ and $\hat{a}_2$ in terms of $\hat{a}_3$ and $\hat{a}_4$:
\begin{equation}\label{e1}
    \hat{a}_2 = \sqrt{\eta t_B} \hat{a}_3 - \iota \sqrt{\eta (1-t_{B})} \hat{a}_4
\end{equation}
and 
\begin{equation}\label{e2}
     \hat{a}_1 =  - \iota \sqrt{ \eta (1-t_{B})} \hat{a}_3 + \sqrt{\eta t_B}\hat{a}_4.
\end{equation}
For the state $\big|0_z\rangle$, at $\tau_0$, we will have vacuum ($|0\rangle$) coming from mode 2 and $|\alpha\rangle$ coming from mode 1. This can be expressed using the displacement operator as $D_1(\alpha) |0 \rangle_3 |0 \rangle_4$, where $D_1(\alpha)= \exp(\alpha \hat{a}_1^\dagger - \alpha^\ast \hat{a}_1 ) $.  The output state at mode 3 is given by $|\sqrt{\eta t_B} \alpha \rangle$.

 In the monitoring line, we have two 50:50 beam splitters and a phase shifter $\phi$ in one arm to create a one-bit delay and two SPAD detectors. Now, in the monitoring line, for the $\ket\alpha \ket \alpha$ as the input state, the output on mode 4 is given as $ \big| \iota\sqrt {1-t_B} \alpha \rangle$ at $\tau_0$. Similarly, for $\tau_1$, the output on mode 4 is $| \iota\sqrt {1-t_B} \alpha \rangle$. Applying the same BS analysis as we have done above, the output state on mode 8 is given as $\big| - \iota \sqrt{\eta}\sqrt{1-t_B}( 1+ e^{\iota\phi}) \alpha/2\rangle$ and mode 9 is given as $\big|\sqrt{\eta}\sqrt{1-t_B}( 1+ e^{\iota\phi}) \alpha/2 \rangle$. 
Therefore, using equations (\ref{e1}) and (\ref{e2}), the gain $G_{\alpha \alpha}^{M_0}$  (Eq. \ref{gaineqn}) is calculated as :
\begin{equation}
    G_{\alpha \alpha}^{M_0} =  (1-p_d)^3 [1-(1-p_d)e^{-\mu \eta (1-t_B)(1+\cos{\phi})/2}] e^{- t_B \mu \eta},
    \label{gainalpha}
\end{equation}
where, $ G_{\alpha \alpha}^{M_0}$ represents the probability that the detector $D_{M_0}$ registers a click, while detectors $D_{M_1}$ and the data line detector do not. For $\phi = \pi/2$, Eq. \ref{gainalpha} becomes:
\begin{equation}
G_{\alpha \alpha}^{M_0} =  (1-p_d)^3 [1-(1-p_d)e^{-\mu \eta (1-t_B)/2}] e^{- t_B \mu \eta}.
\end{equation}
A factor of 2 appears in the exponential of the last term, which arises from the use of an optical switch in the AMZ interferometer, which does not attenuate the pulse, for a higher key rate. Since we have employed 50:50 beam splitters (DLI) in our setup, this factor would be absent.
Similarly, the gain $G_{\alpha \alpha}^{M_1}$ represents the probability that a detection event occurs at the monitoring-line detector $D_{M_1}$, conditioned on the absence of clicks at the other monitoring detector $D_{M_0}$ and the data-line detector $D_T$. To account for false detection events originating from detector dark counts, the dark count probability $p_d$ is explicitly incorporated into the gain expression. The expression of $G_{\alpha \alpha}^{M_1}$ is given as:
\begin{equation}
    G_{\alpha \alpha}^{M_1}= p_d (1-p_d)^3 e^{-2\mu (1-t_B) \eta} e^{-t_B \mu\eta}. 
\end{equation}
For the vacuum decoy state, the gain $G_{00}^{M_i}$ ( where $i=0,1$) arises solely from detector dark counts, as no genuine detection events are expected. Accounting for false clicks via the dark count probability $p_d$ one obtains,
\begin{equation}
    G_{00}^{M_0}= p_d (1-p_d)^3= G_{00}^{M_1}
\end{equation}
These expressions are derived under passive basis choice and coincide with those in \cite{gao_2022}.\\

The bit error rate $\mathcal{E}_Z$ can be calculated from the data line gains directly as: 
\begin{equation}
    \mathcal{E}_Z= \frac{G_{0_z}^{\tau_1}+G_{1_z}^{\tau_0}}{G_{0_z}^{\tau_0}+G_{0_z}^{\tau_1}+G_{1_z}^{\tau_0}+G_{1_z}^{\tau_1}}.
    \label{qber}
\end{equation}
These gains can be observed directly from the data line detector. To account for false detection events originating from dark counts, including those arising from the monitoring line in the passive basis-choice configuration, we derive analytical expressions for the data-line gains using a beam-splitter model as described above. The resulting gains are given by:  $G_{0_z}^{\tau_0} =  (1-p_d)^3[ 1- e^{-\eta t_B|\alpha|^2}]$;
$G_{0_z}^{\tau_1} = p_d (1-p_d)^2$;
$G_{1_z}^{\tau_0} = p_d (1-p_d)^2$;
$G_{1_z}^{\tau_1} = (1-p_d)^3[ 1- e^{-\eta t_B|\alpha|^2}].$\\

For the phase error rate calculation in the finite key regime, we use the method described in \cite{li2024finite}.  A virtual entanglement-based protocol is used to obtain the secure key rate in the finite regime. The security proof provided in \cite{li2024finite} used the quantum leftover hash lemma and the entropic uncertainty relation to derive a formula for the secure key rate in the finite key regime. To deal with correlated random variables, they have used Kato's inequality to estimate statistical fluctuations and ensure security against coherent attacks. Alice prepares her quantum states in the Z-basis as $|0_z\rangle = |0\rangle |\alpha\rangle$ and $|1_z\rangle =  |\alpha\rangle |0\rangle$. Then, the states in the X-basis can be expressed as, 
\begin{equation*}
    |0_x\rangle = \frac{1}{\sqrt{N^+}}( |0_z\rangle+|1_Z\rangle),
\end{equation*} and,
\begin{equation*}
    |1_x\rangle = \frac{1}{\sqrt{N^+}}( |0_z\rangle-|1_Z\rangle),
\end{equation*}
where $N^+= 2(1+e^{-\mu})$, $N^-= 2(1-e^{-\mu})$ and $\mu= |\alpha|^2$ is the mean photon number. 
Alice randomly selects in which basis (X or Z) to prepare her quantum states. If she choses X-basis then she prepares $|0_x\rangle$ and $|1_x\rangle$ with probabilities $N^+/4$ and $N^-/4$. And, if she selects Z-basis then she prepares $|0_z\rangle$ and $|1_z\rangle$ with probability $1/2$.
The optical pulses are then transmitted by Alice to Bob, who employs the identical experimental configuration as the earlier coherent COW-QKD technique. Pulses are monitored in the appropriate basis using a beam splitter for passive basis selection. Both the X and Z bases have the same density matrix. Furthermore, the phase error rate in the COW-QKD protocol is equivalent to the bit error rate in the X-basis, as the density matrix remains identical in both the Z and X bases. This equivalence significantly complicates an adversary's ability to determine whether the prepare-and-measure protocol or the original COW-QKD protocol is being implemented by Alice and Bob. In practical implementations of COW-QKD, the phase error rate corresponds directly to the bit error rate of the X-basis in the prepare-and-measure protocol. The expression for bit error rate of the X-axis is then provided by \cite{li2024finite},\\
\begin{equation}
    \mathcal{E}_x = \frac{N^+( G_{0_x}^{M_1}- G_{0_x}^{M_0})+2(G_{0_z}^{M_0}+G_{1_z}^{M_0}) }  {2(G_{0_z}^{M_0}+G_{0_z}^{M_1}+G_{1_z}^{M_0}+G_{1_z}^{M_1})}
    \label{phaseerrorrate}
\end{equation}

In the experimental COW-QKD setup, the states $|0_x\rangle $ and $|1_x\rangle $ cannot be sent, and hence we cannot calculate the gains $G_{0_x}^{M_i}= \langle 0_x \big|\hat{K}_i^\dagger \hat{K}_i \big|0_x \rangle$ directly.  The decoy states $\big|0\rangle\big|0\rangle$ and $\big|\alpha \rangle\big|\alpha\rangle$ are used to express the monitoring line gains in the X-basis. In view of finite-key effects, establishing security in the finite-key regime is essential for practical deployment of the protocol. This requires accounting for statistical fluctuations between observed and expected quantities and deriving a rigorous lower bound on the final secret key length using appropriate concentration inequalities \cite{kato2020concentration,azuma1967weighted}. Recently, Kato's concentration inequality has shown promise in tightening the estimate, allowing a higher key to be obtained \cite{kato2020concentration}. Using the Kato's concentration inequalities, the upper bounds on corresponding expected values are as follows:
\begin{equation}
    n_y^{M_i^*} \leq \bar{n}_y^{M_i^*} = n_y^{M_i}+\delta_y^{M_i},
    \label{upperbound}
\end{equation}
where, $y= 00, \alpha \alpha $ and $i=0,1$. 
The quantities, $n_{00}^{M_i} $ and $n_{\alpha \alpha}^{M_i}$, represent the total number of clicks obtained in the monitoring line detectors, $D_{M_0}$ and $D_{M_1}$ when decoy pulses $00$, $\alpha \alpha$ are sent, these parameters can be observed directly on the monitoring line detectors.
The statistical fluctuation parameter $\delta_y^{M_i}$ are obtained in the way presented in \cite{li2024finite}. Similarly, the two lower bounds $\underline{n}_y^{M_0^*}$ where, $y= 00, \alpha\alpha$ are calculated as:
\begin{equation}
n_y^{M_0^*} \geq \underline{n}_{y}^{M_0^*} = n_y^{M_0}-\delta_y^{M_0'}  ,
\label{lowerbound}
\end{equation}
There will be 4 upper bounds and 2 lower bounds, and the failure probability for estimating each of the 6 bounds will be considered to be $\epsilon_1$. 
Based on Eq. (\ref{upperbound}) and (\ref{lowerbound}), the upper and lower bounds on the gains of each event will be calculated as:

\begin{equation}
    \overline{G_{y}^{M_i^*}} =  \overline{n_y^{M_i^*}}/n_y ,
    \label{upperboundgain}
\end{equation}
and,
\begin{equation}
    \underline{G_{y}^{M_i^*}} = \underline{n_y^{M_i^*}}/n_y. 
    \label{lowerboundgain}
\end{equation}
By applying these Eqs. (\ref{upperboundgain}) and (\ref{lowerboundgain}), the expected values of the gains of the monitoring line in the X-basis are:
\begin{equation}
\begin{aligned}
\overline{G_{0_x}^{M_1^*}}
&= \frac{1}{N^+}
\left(
  e^{\mu/2}\sqrt{\overline{G_{\alpha\alpha}^{M_1^*}}}
  + e^{-\mu/2}\sqrt{\overline{G_{00}^{M_1^*}}}
\right)^2  \\
&\quad + \frac{N^-}{N^+}
\left(
  \frac{e^{\mu} N^-}{4}
  + e^{\mu}\sqrt{\overline{G_{\alpha\alpha}^{M_1^*}}}
  + \sqrt{\overline{G_{00}^{M_1^*}}}
\right),
\end{aligned}
\end{equation}

\begin{equation}
\begin{aligned}
\underline{G_{0_x}^{M_0^*}}
&= \frac{1}{N^+}
\Big(
  e^{\mu}\,\underline{G_{\alpha\alpha}^{M_0^*}}
  + e^{-\mu}\,\underline{G_{00}^{M_0^*}}
  - 2 \sqrt{
      \overline{G_{00}^{M_0^*}}\,
      \overline{G_{00}^{M_0^*}}
    }
\Big)  \\
&\quad
- \frac{N^-}{N^+}
\Big(
  e^{\mu}\sqrt{\overline{G_{\alpha\alpha}^{M_0^*}}}
  + \sqrt{\overline{G_{00}^{M_0^*}}}
\Big).
\end{aligned}
\end{equation}
On applying these upper and lower bound gains in Eq. \ref{phaseerrorrate}, the expected upper bound on the phase error rate is obtained as \cite{li2024finite}:
\begin{equation}
  \overline{ \mathcal{E}_p^*}= \overline{ \mathcal{E}_x^*} = \frac{N^+\big(\overline{G_{0_x}^{M_1^*}}- \underline{G_{0_x}^{M_0^*}} \big) + 2\big( G_{0_z}^{M_0} + G_{1_z}^{M_0}\big)}{2 \big(G_{0_z}^{M_0} + G_{0_z}^{M_1}+G_{1_z}^{M_0}+G_{1_z}^{M_1}\big) } .
  \label{boundonphaseerrorrate}
\end{equation}
To calculate the bound on phase error rate in the observed value case, Kato's inequality is used again, and the expected number of clicks corresponding to phase errors is provided by $\overline{n_p^*} = N \times  \overline{ \mathcal{E}_p^*} $. Therefore, the upper bound on the observed value using Kato's inequality is given as:
\begin{equation}
    n_p \leq \overline{n_p}= \overline{n_p^*} +\delta_p , 
\end{equation}

where, $\delta_p= \sqrt{\frac{1}{2}n_z \ln\epsilon_2^{-1}}$; $\epsilon_2$ is the failure probability for estimating $\overline{n_p}$. 
The upper bound on the phase error rate is then given by:
\begin{equation}
    \overline{\mathcal{E}_p}= \overline{n_p}/n_z . 
    \label{actualPER}
\end{equation}

Using the finite key analysis, the secure key length for the COW-QKD system can be written as \cite{li2024finite}

\begin{equation}
    \mathcal{R} \geq n_z [ 1- h( \overline{\mathcal{E}_p})]-leak_{EC} - \log_2 \big( 2/ \epsilon_{cor}\big)- 2 \log_2 \big(5/\epsilon_{sec} \big).
    \label{keyrate}
\end{equation}
Here, $h(.)$ represents the Shannon entropy. The parameters used in Eq. \ref{keyrate} are provided in the next section.

\section{Experimental Results}\label{experiment}

We have used a two-state decoy variant of the COW-QKD in our lab, in which we have used a $1550.12$ nm PS-NLL laser with pulse frequency as $1$ $\rm{GHz}$. The repetition rate of the incoming pulse (logical qubit) is then $500$ $\rm{MHz} = 5 \times 10^8$ Hz which therefore leads $N= 5 \times 10^8$ rounds per second. The total counts correspond to the number of pulses successfully detected at Bob’s side and are strongly influenced by the quality of the optical channel, the source parameters, and the detector characteristics. The average number of photons per pulse in our experiment is $\mu=0.5$. Alice and Bob are connected with each other via an optical fiber of length $80$ km with attenuation coefficient of  $0.2$ $\rm{dB/km}$. So there is a net loss $16$ $\rm{dB}$ from Alice to Bob.  The gap between two pulses is $1  \rm{ns}$. The dark count rate of the detector is $900\, \rm{counts/sec}$. So, the dark count rate probability is $p_d= \frac{900 \,\rm{counts/sec}}{500 \rm{MHz}} = 1.8 \times 10^{-6}$. The transmittance $t_B$ of the $90:10$ beam splitter is $0.90$. The transmittance of the optical fiber with length $l$ is $10^{-0.02 l} $. Considering detector efficiency also into account, the total loss is given by, $\eta= \eta_d \times 10^{0.02 l}$ and here we have modelled this loss in terms of the amplitude damping channel between Alice and Bob. For the COW-QKD setup, the parameters are specified in Tables \ref{cow_table_1} and \ref{table:1}. For finite key analysis, the security bounds of correctness and security are fixed to be $\epsilon_{cor} = 10^{-15}$ and $\epsilon_{sec} = 10^{-10}$. 

\begin{table*}[t]
\centering
\caption{Specifications of the components used in the COW-QKD setup}

\setlength{\tabcolsep}{14pt} 

\begin{tabular}{l l}
\hline\hline
Component / Technique & Property / Type \\
\hline
Laser & 1550.12 nm PS-NLL laser (Teraxion) \\
Fiber & SMF-28 (ITU-T G.652D), loss = 0.2 dB/km \\
Phase modulator & MPZ-LN series \\
Intensity modulator & MXER\_LN\_10 \\
Random number generator & Quantum random number generator (QRNG) \\
Operating temperature & $10^\circ\mathrm{C}$ -- $28^\circ\mathrm{C}$ \\
Detector & SPD\_OEM\_NIR (Aurea Technology) \\
Error correction & LDPC \\
Privacy amplification & Toeplitz-based hashing \\
\hline\hline
\end{tabular}

\label{cow_table_1}
\end{table*}
\begin{table}[h!]
\centering
\caption{Parameters which are affecting the key rate of COW protocol}
\begin{tabular}{||c c ||} 
 \hline
 Parameters  &  COW \\ [0.5ex] 
 \hline\hline
 Average Photon Number &  0.5 \\ 
 \hline
 Loss in Fibre ($l_{f}$) &  0.2 dB/km \\
 \hline
 MZI Loss ($l_{m}$) &  2 dB \\
 \hline
 Distance (d)  &  80 km  \\
 \hline
 Pulse frequency (f) &  01 GHz\\
 \hline
 Detector efficiency ($\eta$) &  10-25\%\\
 \hline 
 Dead time ( $t_d$ ) &  30-50 $\mu$ s \\
 
\hline
 Disclose rate (DR)   &  8-15\% \\
 \hline
 Compression ratio (CR)   &  50-90\%  \\ [1ex]
 \hline
\end{tabular}
\label{table:1}
\end{table}

\begin{figure}
    \centering
    \includegraphics[width=0.6\linewidth]{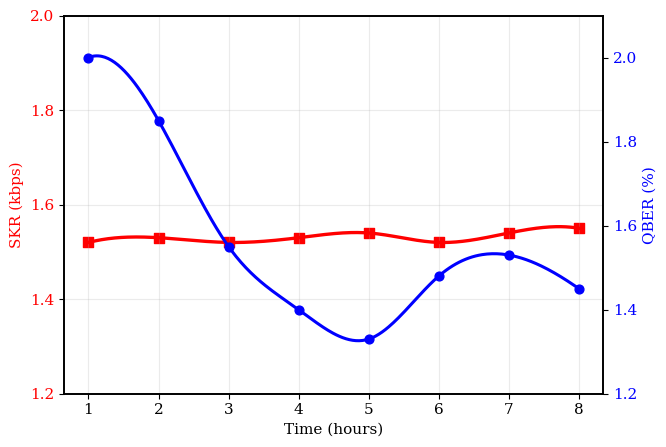}
    \caption{(Color online) Stability Analysis of QBER(\%) and SKR(Kbps) w.r.t time (hrs) (at $\eta_d = 0.20$, $DT= 30 \mu s$, $DR=10\%$, $CR=80\%$)}
    \label{fig:stabilitycheck}
\end{figure}

\begin{figure}[htbp]
\centering
\includegraphics[width=0.6\linewidth]{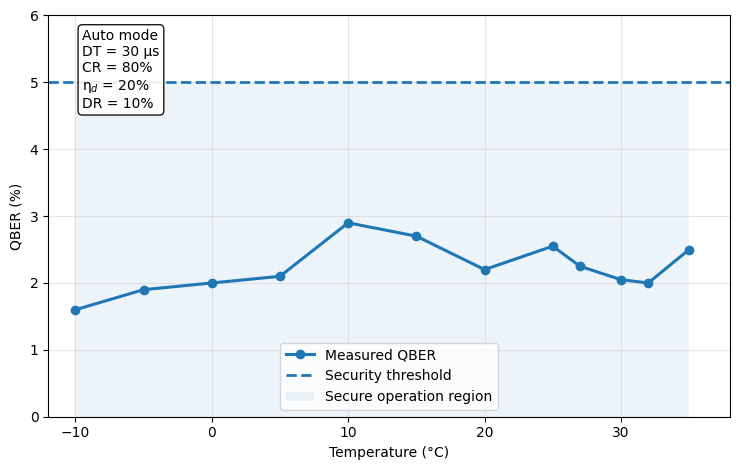}
\caption{ (Color online) Variation of QBER vs temperature (at $\eta_d = 0.20$, $DT= 30 \mu s$, $DR=10\%$, $CR=80\%$) }
\label{Qber_temp}
\end{figure}

\begin{figure}[htbp]
\centering\includegraphics[width=0.6\linewidth]{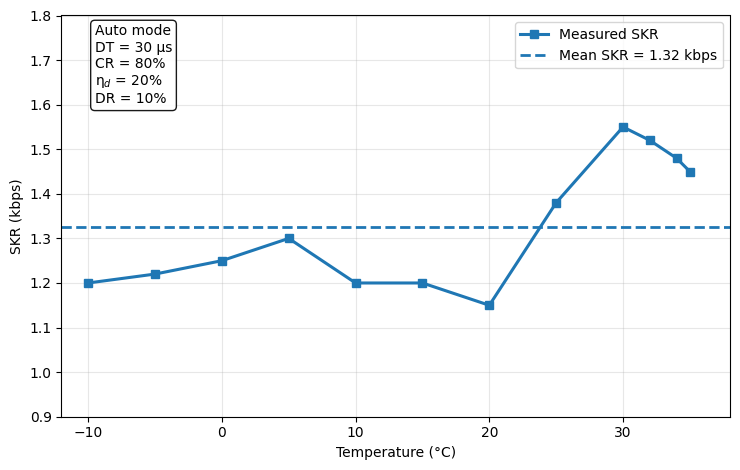}
\caption{(Color online) Variation of secure key rate vs temperature (at $\eta_d = 0.20$, $DT= 30 \mu s$, $DR=10\%$, $CR=80\%$)}
\label{kr_temp}
\end{figure}

 With the specifications mentioned in Table \ref{table:1},  we ran the COW-QKD setup for several hours and kept the optical fiber in environmental test chamber to obverse the QBER and secure key rate with respect to various realistic scenarios. Fig. \ref{fig:stabilitycheck} shows the stability analysis of secure key rate and QBER with respect to the continuous 8 hours of operation. We can see that we are getting a stable key rate of around 1.6kbps during the entire duration of runtime. As far as the QBER is concerned, we can see some minor fluctuations, but after few hours of run time, it has stabilized to around 1.6 percent. So we can clearly see that the experimentally realized COW-QKD is producing stable secure key rates. Then, we put the fiber link between Alice and Bob in the environmental test chamber to see if there is any appreciable changes in the secure key rates with respect to the realistic operating temperatures. Fig. \ref{Qber_temp} shows that QBER has remained stable (between 2-3 percent) and well below the threshold value of 5 percent in the operational temperature zones. Fig. \ref{kr_temp} shows that key rate is also stable (between 1.2-1.6 kbps) in the operational temperature zones. Next, we will use the experimental parameters of our physical system and finite key analysis to find the distance upper bounds up-to which the secure key rate can be generated between Alice and Bob.
 

\begin{figure}[htbp]
   \centering
  \includegraphics[width=0.9\linewidth]{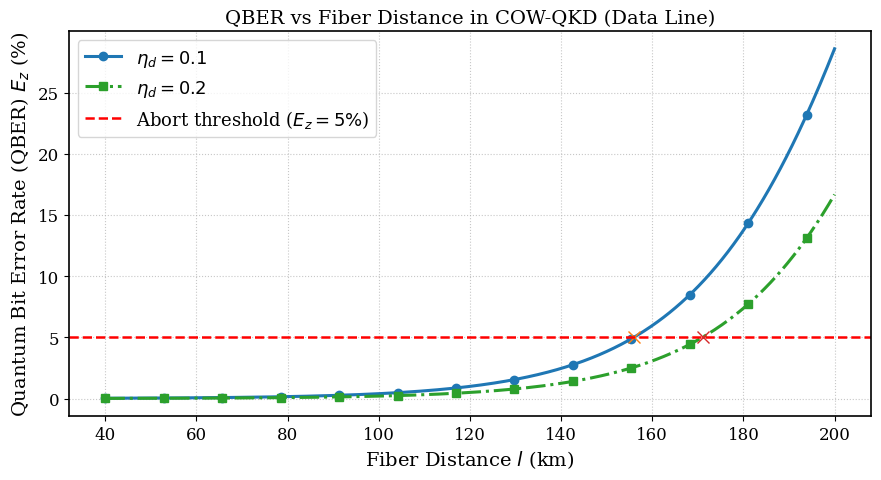}
 \caption{(Color Online) Variation of QBER (\%) vs Fiber Distance (km). The blue line and green line show the variation when the detector efficiency $\eta_d = 0.1$, and $\eta_d = 0.2$, respectively. The other parameters are $t_B=0.90$ and  $\mu=0.5$. }
\label{fig:qbervsdist}
\end{figure}

Using our experimental parameters, we have tried to model QBER, phase error rate, and secure key rate for the finite key regime. Quantum bit error rate for our COW-QKD system has been found to be given by Eq. \ref{qber}. We have now tried to find out how QBER scales up with distance by considering our experimental parameters. We have considered 5 percent QBER as the threshold beyond which the protocol will be aborted.  Fig. \ref{fig:qbervsdist} illustrates the variation of the Quantum Bit Error Rate (QBER)  $E_z$ with fiber length in a COW-QKD setup. As the transmission distance increases, the QBER also rises due to the combined effects of increasing channel loss and constant detector dark counts. The red dashed line marks the abort threshold at 5\%, which represents the maximum tolerable error rate for secure key generation. The intersection point, highlighted by a red dot, occurs at $~\approx$ 156 km (when detector efficiency $\eta_d=0.1$ ), and $~\approx$ 171 km (when detector efficiency $\eta_d=0.2$). This indicates the effective operational limit of the protocol under the given parameters. Beyond this distance, the error rate becomes too high for secure quantum communication, and the protocol must be aborted. Here, in our lab, we have used telecom-grade fiber that has a loss of 0.2 dB/km, and the detector efficiency of SPAD is also low. If we could replace the telecom-grade optical fiber with ultra-low-loss optical fiber (fiber loss 0.15 dB/km) and used  SSNPAD ( $\eta_d = 0.95$) instead of SPAD, then the upper bound on distance is found to be  273.14 km. 
 
Now we will estimate the phase error for our system. The number of bits that are revealed in the error-correction step $leak_{EC}$ is $fn_z h (\mathcal{E}_z)$, where the correction efficiency is set to be $f=1.1$ and $n_z$ is the total number of clicks received on the data line. For a distance of $l=100 \rm{km} $, at the detector efficiency $\eta_d = 0.1$ and dead time of the detector, $DT= 50 \mu s$, the total number of clicks observed on the data line is $n_z= 14811 \rm{\ s}$. And for $l=100 \rm{km}$, the detector efficiency $\eta_d = 0.2$ and dead time of the detector, $DT= 30 \mu s$ the total number of clicks observed on the data line $n_z= 26460 \rm{\ s}$. The number of states $|\alpha \rangle |\alpha \rangle $ sent by Alice is $n_{\alpha\alpha}= N\times \mathrm{p}_{d_1}$ and the number of states $|0\rangle|0\rangle$ sent by Alice is $n_{00}= N\times \mathrm{p}_{d_2}$.
In the monitoring line, $n_{\alpha \alpha}^{D_{M_0}}$ and $n_{\alpha \alpha}^{D_{M_1}}$  are the total number of clicks observed on the detector $D_{M_0}$ and $D_{M_1}$ for the pulse $|\alpha\rangle |\alpha\rangle$. For the decoy state $|0\rangle|0\rangle$, since this is a vacuum pulse, any contribution must be due to the dark counts. 
\begin{figure}[htbp]
   \centering
    \includegraphics[width=0.6\linewidth]{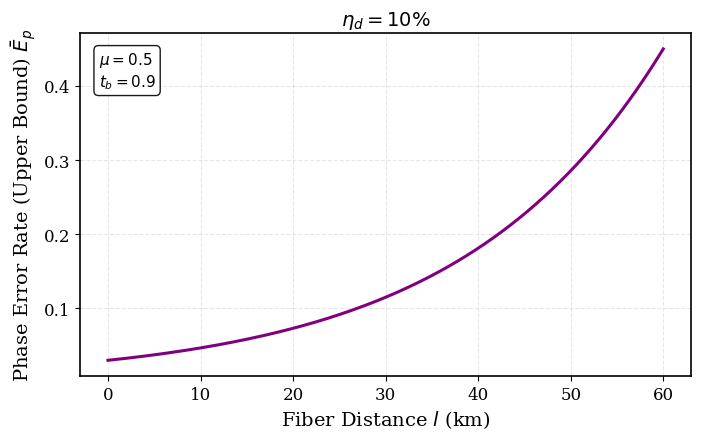}
    \caption{(Color Online) Variation of Phase Error Rate (Upper Bound) w.r.t. Fiber Distance at $\eta_d=0.1$ and $DT=50 \mu s$}
   \label{fig:PERVSdistance1}
\end{figure}
\begin{figure}[htbp]
    \centering
    \includegraphics[width=0.6\linewidth]{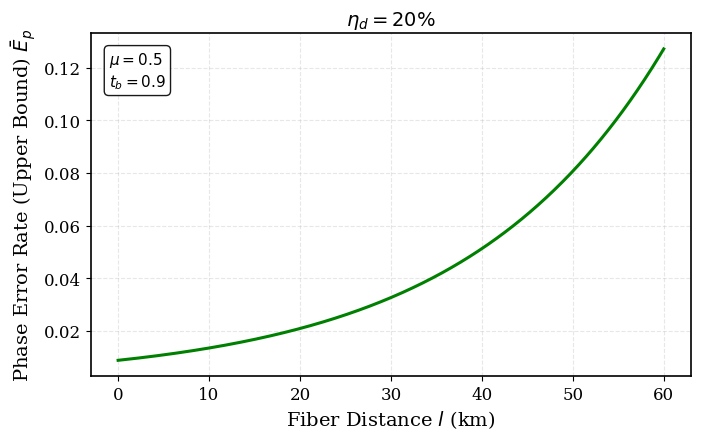}
    \caption{ (Color Online) Variation of Phase Error Rate (Upper Bound) w.r.t. Fiber Distance at $\eta_d=0.2$ and $DT=30 \mu s$}
    \label{fig:PERVSdistance2}
\end{figure}

As explained in the above section, the phase error rate in the finite regime cannot be calculated directly. Using Kato's inequality, the upper bounds on the expected and observed values of the phase error rate are calculated. The probability of failure $\epsilon_1$ and $\epsilon_2$ are set to be $10^{-11}$, as $\epsilon_1 = \epsilon_2 = \epsilon_{sec}/10$ for the finite key regime. Using Eq. \ref{actualPER}, we can calculate the upper bound on the phase error rate.
Fig. \ref{fig:PERVSdistance1} and \ref{fig:PERVSdistance2}, shows the variation of phase error rate $\overline{\mathcal{E}_p}$ (upper bound) with respect to the fiber distance $l$( km) at different detector efficiencies ($\eta_d$) and dead times ($DT$). 
As the fiber distance increases, the upper bound on the phase error rate $\overline{\mathcal{E}_p}$ increases monotonically in both cases. This trend is primarily due to channel loss, which grows with distance and reduces the probability of detecting signal photons. As a result, dark counts and finite-size statistical fluctuations contribute more significantly to the observed detection events.  When a higher detector efficiency $\eta_d=0.20$ is employed, the phase error rate remains noticeably lower over the entire distance range. Improved detection efficiency enhances the effective signal-to-noise ratio, enabling more reliable estimation of coherence-related parameters and mitigating the impact of loss-induced fluctuations. In contrast, for a lower detector efficiency ($\eta_d=0.1$), the phase error bound increases much more rapidly with distance, ultimately imposing a stronger constraint on secure key generation at longer transmission lengths. 

\begin{figure}[htbp]
    \centering
\includegraphics[width=0.6\linewidth]{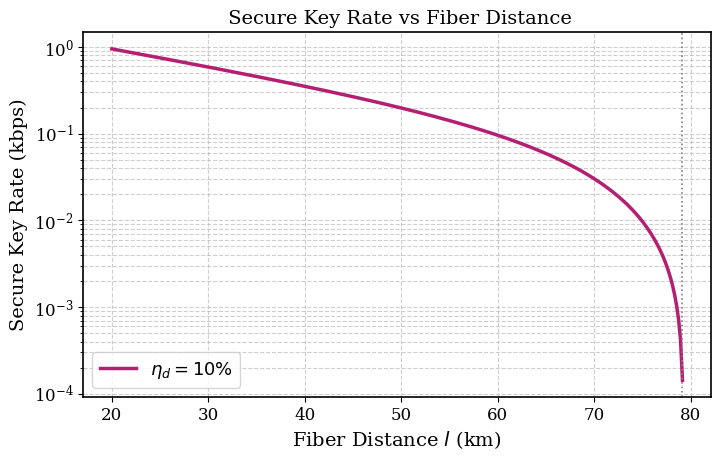}
     \caption{(Color Online) Variation of Secure Key Rate vs Fiber Distance $l$ (km) at total number of rounds $N= 5\times 10^8$, detector efficiency $\eta_d = 0.1\%$ and dead time $DT=50 \mu s$.}
    \label{fig:SKRatDE0.1}
\end{figure}
\begin{figure}
    \centering
    \includegraphics[width=0.6\linewidth]{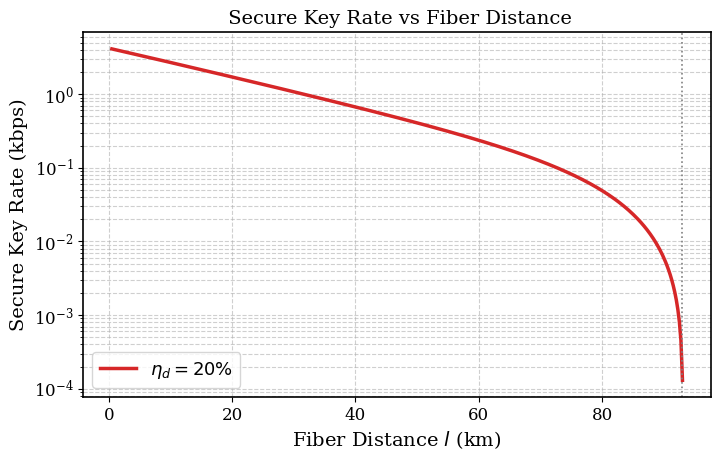}
    \caption{(Color Online) Variation of Secure Key Rate vs Fiber Distance (km) at total number of rounds $N= 5\times 10^8$, detector efficiency $\eta_d = 0.2\%$ and dead time $DT=30 \mu s$.}
    \label{fig:skrvsdist0.2}
\end{figure}
The secure key-length can be calculated using Eq. (\ref{keyrate}). Fig. \ref{fig:SKRatDE0.1} shows the variation of secure key rate (kbps) w.r.t the fiber distance ($l$) at detector efficiency $\eta_d=0.1$, dead time $DT= 50 \rm{\mu s} $. 
At shorter distances $20-40$ km, the key rate is relatively high $\approx 1 \rm{Kbps}$, because the channel losses are small and statistical fluctuations are minimal. As we increase the fiber distance, attenuation reduces the signal detection probability exponentially. The finite-key corrections significantly reduce the extractable key length. Beyond $75-80$ km, the key rate drops sharply to zero. This sharp cut-off indicates the maximum usable transmission distance in the finite regime for the considered experimental parameters. Fig. \ref{fig:skrvsdist0.2} shows the variation of secure key rate (kbps) w.r.t. the fiber distance ($l$) at detector efficiency $\eta_d=0.2$, dead time $DT= 30 \rm{\mu s} $. We observed that the secure key rate decreases monotonically with increasing fiber distance, and a sharp cut-off is observed near $\approx  90$ km. SKR remains in the Kbps regime till $\approx 60-65  $ km. Since the secure key rate depends on losses in the optical fiber and efficiency of SPAD, so this distance can be increased if we replace telecom fiber with ultra low loss optical fiber and low-efficiency SPADs with more efficient superconducting nanowire single photon detectors (SNSPDs). 

\section{Conclusion} \label{conclusion}
In this work, finite key analysis of the COW-QKD protocol is performed to obtain realistic security bounds. Specifically, a clean and easy-to-use framework for performing such an analysis, which can be used by others to perform similar analyses, is provided by extending the existing works. The construction of this framework was not trivial because of its reliance on coherence measurements and correlated detection events, which make rigorous parameter estimation
essential for realistic security claims.
In addition, experimental realisation of COW-QKD system is reported here and the finite key analysis of the experimentally realized COW-QKD system is performed with specific attention to  the analysis of the parameters such as QBER, phase error rate and secure key rate under realistic parameters.  COW-QKD system has been realized by many earlier research groups including some of the present authors because of the relatively less challenging implementation. Here,  also  its implemented but in contrast to the earlier works, a finite key analysis is performed to obtain a realistic estimate of the distance up-to which this particular implementation of QKD can be considered to be secure. It is observed that the secure key rate depends on a number of factors such as beam-splitting ratio, attenuation coefficient of optical fiber, detector efficiency, detector dead time. We modelled the COW-QKD system by considering the amplitude damping channel between the sender and the receiver. We have been able to show that COW-QKD system can be considered to be secure over a distance in the range of $150$ km if we use normal telecom-grade optical fiber and SPAD. Specifically, it is secure up-to $~156$ ($~171$) km of optical fiber with 0.2 dB loss per km if detector efficiency is 0.1 (0.2) and other parameters are considered to be the same as those used in the experimental realization reported here. The distance for secure key generation can be increased to $~273$ km if we use ultra low loss optical fiber and highly efficient SNSPD. We conclude this work with the hope that the clarity provided in the present work will lead to the desiging of improved optical components and detectors leading to longer distance over which COW-QKD can be performed as it clearly illustrates the role of different experimental parameters on the bounds obtained here. Furhter, the present study is restricted to COW-QKD protocol which is an specific example DPR protocols, we hope that the rationale used here will lead to easy-to-use frameworks for finite key analysis of DPS and other QKD protocols of DPR family.

\section{Acknowledgment}
The authors would like to acknowledge Mr. Jimmy Tamakuwala and Mr. Kamran Nazir for their help in this work and DRDO, India, for providing the financial support.

\section{Disclosures}
The authors declare no conflicts of interest.

\section{Data availability.}
 Data underlying the results presented in this paper are not publicly available at this time but may be obtained from the authors upon reasonable request.


\bibliography{FiniteV1}

\end{document}